\documentclass[a4paper,11pt]{article}
\pdfoutput=1 % if your are submitting a pdflatex (i.e. if you have
\usepackage{jcappub} % for details on the use of the package, please
\usepackage[T1]{fontenc} % if needed
\usepackage{booktabs}
\usepackage{graphicx}

\usepackage{booktabs}
\usepackage{float}

\usepackage{siunitx}
\usepackage{placeins}
\usepackage{graphicx}
\usepackage{subcaption}
\usepackage{amsmath}
\usepackage{comment} 
\usepackage{amssymb}
\usepackage{hyperref}
\usepackage{lmodern}
\usepackage[dvipsnames]{xcolor}
\usepackage{pgfgantt}

\title{\boldmath Bounds on the Tsallis Parameter from a deformed Neutrino Sector in the Early Universe}

 \author{Matias P. Gonzalez}
\affiliation{Departamento de Física, Universidad Católica del Norte,\\Avenida Angamos 0610, Chile}

\emailAdd{matias.gonzalez03@alumnos.ucn.cl}

\abstract{We generalize neutrino energy density content in the early universe near BBN era $T\simeq1$ MeV within Tsallis nonextensive statistics. By using Curado-Tsallis constraints we obtain generalized distribution functions $f_q(E)$. We compute the generalized thermodynamic integral for the energy density $\rho_q$. We define a reescaling $R^{(\xi = +1)}_{\rho}(q) = \rho_q/\rho^{\rm std}$ which is a ratio between the deformed energy density and the standard extensive case. The last was used to directly map and deform neutrino content via the effective number of neutrinos $N_{\rm eff}$. The deformation prediction was confronted against CMB$+$BAO and BBN data for $N_{\rm eff}$ by a joint/combined $\chi^2$ type-fit. We obtained the constraints
$|q-1|\le 1.09\times 10^{-2}$ (95\% CL) and $|q-1|\le 1.32\times 10^{-2}$ (99\% CL) from the combined analysis by numerically calculating the best value of the Tsallis parameter $q_{\rm best}$.}

\begin{document}
\maketitle
\flushbottom

\section{Introduction}
\label{sec:intro}
The standard thermal history of the early Universe has been remarkably successful in accounting for a wide range of observations, and it provides a consistent framework to reconstruct the evolution of the cosmic plasma back to very early times. In this picture, the primordial bath at high temperatures is well described by Boltzmann-Gibbs (BG) statistical mechanics: particle species remain in thermal equilibrium as long as their interaction rates exceed the Hubble expansion rate, and progressively decouple as the Universe expands and cools \cite{KolbTurner1990,Dodelson2003}.

In the MeV era, relevant for neutrino decoupling and the onset of big-bang nucleosynthesis (BBN), the relativistic energy density is dominated by photons and neutrinos \cite{KolbTurner1990,Dodelson2003}. After neutrino decoupling, the subsequent electron-positron annihilation reheats the photon bath, leaving neutrinos cooler than photons and fixing the standard relation between the two temperatures in the instantaneous-decoupling approximation \cite{Bennett:2021NeffII}. A convenient way to quantify the relativistic content during and after this epoch is through the effective number of neutrino species, $N_{\rm eff}$, which parametrizes the neutrino (and any additional) contribution to the total radiation energy density relative to photons \cite{Planck2018CosmoParams}.

Since $N_{\rm eff}$ is directly sensitive to changes in the neutrino energy density, it provides a clean observational handle on departures from the standard thermal picture. In particular, non-standard distortions of the neutrino phase-space distribution (arising from new interactions, residual out of equilibrium effects, or more general deformations of the underlying statistical framework) can be mapped into shifts in $N_{\rm eff}$, making it a powerful probe of early-Universe physics \cite{CyburtFieldsOliveYeh2016}.

To model a controlled departure from the standard thermal picture, we consider a deformation of the underlying statistical framework within nonextensive Tsallis statistics \cite{Tsallis1988,CuradoTsallis1991,TsallisMendesPlastino1998}. In this approach, generalized equilibrium distribution functions follow from the maximum-entropy principle and depend on a single nonextensivity parameter, $q$, which effectively captures deviations from BG behavior \cite{Tsallis1988,CuradoTsallis1991}. Such deviations can be interpreted as a phenomenological imprint of long-range interactions, strong correlations, memory effects, or non-thermal tails in the phase-space distribution \cite{Tsallis2009Book}.

In this work we apply the Tsallis deformation to the neutrino sector and quantify its impact on the relativistic energy budget through the effective number of neutrino species, $N_{\rm eff}$. By computing the Tsallis-modified neutrino energy density, we obtain a direct mapping $q \mapsto \Delta N_{\rm eff}$, which enables a straightforward inference of bounds on $q$ from current determinations of $N_{\rm eff}$. In particular we use CMB$+$BAO and BBN data \cite{Planck2018CosmoParams,CyburtFieldsOliveYeh2016}.

In the literature independent analyses have derived bounds on departures from extensivity in closely related early-Universe settings, consistently indicating that viable scenarios require $q$ to remain close to unity. For instance, Ghoshal and Lambiase confront Tsallis-based cosmological deformations with BBN and CMB observables and obtain limits that confine $q$ to percent-level deviations from the BG value \cite{ghoshal2021constraintstsalliscosmologybig21}. Along complementary lines, T{\i}rnakl{\i}, B{\"u}y{\"u}kk{\i}l{\i}{\c c} and Demirhan use generalized quantum distributions and a Tsallis-modified Planck law to place bounds on $q$, likewise finding that only small departures from $q=1$ are allowed \cite{Tirnakli_1998}. Together, these results reinforce the expectation that any nonextensive effects relevant for the early Universe must be tightly constrained around the extensive limit such as in \cite{TorresVucetichPlastino1997}.

The remainder of this paper is organized as follows: in section~\ref{sec:Basiccosmology} we review the standard thermal history around neutrino decoupling and $e^\pm$ annihilation, introduce the thermodynamic integrals from standard theory and summarize the parametrization of the radiation energy density in terms of the effective number of neutrinos $N_{\rm eff}$. In section~\ref{sec:Tsallis} we outline the essentials of Tsallis nonextensive statistics and present the generalized equilibrium distribution functions in the Curado-Tsallis prescription. In section~\ref{sec:Neutrinosectordeform} we implement a neutrino-sector deformation, derive the corresponding rescaling of the neutrino energy density and its mapping to $\Delta N_{\rm eff}(q)$, and describe the inference methodology and datasets used in the analysis. In section~\ref{sec:results} we present the resulting constraints on $q$ from BBN and CMB$+$BAO determinations of $N_{\rm eff}$ and discuss their implications. Finally, in section~\ref{sec:conclusion} we summarize our findings and comment on possible extensions.

\section{Early Universe and macroscopic observables}
\label{sec:Basiccosmology}

In this section we briefly review the standard assumption of thermal equilibrium in the early Universe and the key epochs of interest, with emphasis on neutrino decoupling and the subsequent $e^\pm$ (electron-positron) annihilation that reheats photons. Statistical mechanics provides the bridge between microphysics and macrophysics by relating the phase-space distribution functions to macroscopic quantities such as $n$, $\rho$, and $P$, which ultimately enter the cosmological observables. Finally, we summarize the radiation content around this epoch and introduce a convenient parametrization in terms of the photon energy density which is extremely useful for the main purpose of this paper.

\subsection{Thermal history and the key epoch}
In its earliest stages, the Universe formed a dense and hot plasma that is commonly assumed to be in thermal equilibrium and well described by Boltzmann-Gibbs statistical mechanics. This framework underlies the standard treatment of the evolution of the particle species in the primordial bath. Here we are particularly interested in the epoch around $T\simeq 1~\mathrm{MeV}$ (near the onset of BBN) when electron-positron pairs annihilate and transfer entropy to photons (photon reheating). As a consequence, after neutrino decoupling the photon temperature $T_\gamma$ evolves differently from the neutrino temperature $T_\nu$, the last is described by the standard ratio $T_{\nu}/T_{\gamma} = (4/11)^{1/3}$ (in the instantaneous-decoupling approximation) showing that neutrinos are colder than photons after reheating \cite{KolbTurner1990,Dodelson2003,Bennett:2021NeffII}.

The macroscopic properties of each species in the plasma follow from the equilibrium phase-space distribution function $f(E)$, which gives the mean occupation number of a single-particle state of energy $E$,
\begin{equation}
  f(E) \;=\; \frac{1}{\exp\!\big[(E-\mu)/T\big] + \xi}\,,
  \label{eq:distribucion_estandar}
\end{equation}
where $T$ is the temperature, $\mu$ the chemical potential, and $\xi$ specifies the statistics: $\xi=+1$ for fermions (Fermi-Dirac), $\xi=-1$ for bosons (Bose-Einstein), and $\xi=0$ for the classical Maxwell-Boltzmann limit \cite{KolbTurner1990,Dodelson2003}.

\subsection{Thermodynamic integrals and equation of state}
Equation~\eqref{eq:distribucion_estandar} is the starting point for connecting microscopic physics to macroscopic (and ultimately observable) quantities. Once the distribution function $f(E)$ is specified, the number density $n$, energy density $\rho$, and pressure $P$ follow from standard momentum-space integrals,
\begin{align}
  n \;&=\; g \int \frac{d^3p}{(2\pi)^3}\, f(E)\,, 
  \label{eq:densidad_numerica} \\[0.4ex]
  \rho \;&=\; g \int \frac{d^3p}{(2\pi)^3}\, E(p)\, f(E)\,, 
  \label{eq:densidad_energia} \\[0.4ex]
  P \;&=\; g \int \frac{d^3p}{(2\pi)^3}\, \frac{p^2}{3\,E(p)}\, f(E)\,,
  \label{eq:presion}
\end{align}
where $E(p)=\sqrt{p^2+m^2}$ and $g$ accounts for the internal degrees of freedom (spin, polarization, etc.), assuming they are populated according to thermal equilibrium. A standard caveat applies to neutrinos: in the Standard Model only left-handed neutrino and right-handed antineutrino helicity states are efficiently populated \cite{KolbTurner1990,Dodelson2003}.

Throughout this work we focus on the radiation-dominated era, for which the stress-energy tensor is well approximated by a barotropic equation of state,
\begin{equation}
  P \;=\; \omega\,\rho \,, \qquad \omega \;=\; \frac{1}{3}\,,
\end{equation}
so that $P=\rho/3$ for relativistic species. This relation provides a useful consistency check and allows one to infer the pressure directly from the energy density in the ultra-relativistic limit.

\subsection{Radiation content}
At the temperatures of our interest it is convenient to parametrize the total radiation energy density in terms of photons and neutrinos (after $e^\pm$ annihilation) as
\begin{equation}
  \rho_r \;=\; \rho_{\gamma} + \rho_{\nu} =  \rho_\gamma\bigl(1+k\,N_{\rm eff}\bigr)\,,
  \label{eq:BBNparam}
\end{equation}
where $\rho_\gamma$ is the photon energy density, $\rho_{\nu}$ is the neutrino energy density and $N_{\rm eff}$ denotes the effective number of neutrino species. In the Standard Model the effective number of neutrino species is given by $N_{\rm eff} = 3.0440\pm0.0002$ from neutrino decoupling theory (which includes small corrections beyond instantaneous decoupling). The factor
\begin{equation}
  k \;=\; \frac{7}{8}\left(\frac{4}{11}\right)^{4/3}\label{eq:factor},
\end{equation}
encodes the fermionic nature of neutrinos (the $7/8$ factor) and the standard temperature ratio $T_\nu/T_\gamma=(4/11)^{1/3}$ that results from photon reheating after $e^\pm$ annihilation. As we defined before, Eq.~\eqref{eq:BBNparam} will be useful when we deform the neutrino content by using Tsallis Statistics \cite{Planck2018CosmoParams,Bennett:2021NeffII}.

\section{Tsallis Statistics}
\label{sec:Tsallis}
Tsallis statistics is the main tool we employ to achieve our goal of constraining the nonextensivity Tsallis parameter. In this section we briefly review the formalism, starting from the definition of the Tsallis entropy. We then show how generalized equilibrium distribution functions follow from the maximum-entropy principle when Curado-Tsallis constraints are imposed.

\subsection{Tsallis Entropy and nonextensivity}
The theoretical foundation of Tsallis statistics lies in the concept of non-extensivity. This provides a complete framework for describing complex systems, those with a behavior which  deviates from the standard Boltzmann-Gibbs statistics. Typically, these systems are characterized by the presence of long-range interactions, such as gravity, or by being in non-equilibrium thermodynamic conditions (quasi-equilibrium states). In the early-Universe context, mild deviations from the Boltzmann-Gibbs distributions can translate into measurable shifts in the radiation energy density and hence in $N_{\rm eff}$.

The pillar of Tsallis statistics is the generalization of Boltzmann-Gibbs entropy $S_{\rm BG}$ through a new functional, Tsallis entropy $S_q$ \cite{Tsallis1988,Tsallis2009Book}. This quantity is the starting point for constructing the entire non-extensive framework. Its mathematical definition is given by:
\begin{equation}
    S_q \equiv k\frac{1 - \sum_i p^q_i}{q-1}\label{eq:tsallisentropy},
\end{equation}
where $\{p_i\}$ is the set of probabilities for the system’s microstates, $k$ is a constant analogous to Boltzmann’s, and $q$ is a real number known as the Tsallis parameter. It is essential to note that in the limit $q \to 1$, this expression formally recovers Boltzmann-Gibbs entropy, establishing Tsallis formalism as a generalization of standard statistical mechanics.

While $S_{\rm BG}$ is strictly additive, Tsallis entropy is non-additive (or, more precisely, pseudo-additive). For a system composed of two statistically independent subsystems, A and B, the total entropy of the combined system is not the sum of the individual entropies but obeys the following composition rule:
\begin{equation}
    S_q(A+B) = S_q(A) + S_q(B) + \frac{1-q}{k}S_q(A)S_q(B) \label{eq:pseudo_aditividad}.
\end{equation}
The last term in Eq.~(\ref{eq:pseudo_aditividad}) quantifies the deviation from additivity and is a direct manifestation of the correlations or long-range interactions that the non-extensive formalism seeks to describe. It is clear that in the limit $q \to 1$, standard Boltzmann-Gibbs additivity is recovered, $S_{\rm BG}(A+B) = S_{\rm BG}(A) + S_{\rm BG}(B)$ \cite{Tsallis1988}. This property has two particular cases: sub-additivity $S_q(A+B) < S_q(A)+S_q(B)$, which allows for heavy tails in the velocity distribution implying strong correlations, and super-additivity $S_q(A+B) > S_q(A)+S_q(B)$, which can be associated with an effective cutoff / depleted high-energy tail \cite{Tsallis2009Book}.

Tsallis entropy, $S_q$, shares several fundamental properties with Boltzmann-Gibbs entropy $S_{\rm BG}$, such as non-negativity and the fact that it reaches its maximum value for a uniform probability distribution (equiprobability). However, the crucial distinction between them lies in the property of additivity as seen before.

\subsection{Generalized Distribution Functions}
To extremize $S_q$ under macroscopic constraints, we adopt the Curado-Tsallis (CT) prescription \cite{CuradoTsallis1991,TsallisMendesPlastino1998}, namely the use of \emph{unnormalized} $q$-expectation values for the total energy and particle number,
\begin{align}
\overline{E}=\sum_i p_i^{\,q}E_i,\qquad
\overline{N}=\sum_i p_i^{\,q}N_i,
\end{align}
while keeping the standard probability normalization $\sum_i p_i=1$. This setup yields compact expressions and is particularly convenient for numerically robust implementations in cosmological applications.

Applying the maximum-entropy principle within this framework leads to the Tsallis-deformed single-particle distributions \cite{Buyukilic1995},
\begin{align}
f_{q}(E) \;=\;
\frac{1}{\big[\,1+(q-1)\beta(E-\mu)\,\big]^{\!\frac{1}{q-1}}+\xi}
\;=\;\frac{1}{e_q~\!\big(\beta(E-\mu)\big)+\xi}\,,
\qquad \beta\equiv 1/T,
\label{eq:fq}
\end{align}
where $\xi=-1$ corresponds to Bose-Einstein (BE), $\xi=+1$ to Fermi-Dirac (FD), and $\xi=0$ to Maxwell-Boltzmann (MB) statistics, and $e_q$ denotes the $q$-exponential \cite{Tsallis2009Book}. In the remainder, this parametrization provides the central ingredient to implement controlled deformations of the neutrino sector and, consequently, of the relativistic energy budget of the Universe.

\section{Deforming the neutrino sector}
\label{sec:Neutrinosectordeform}

Having reviewed the standard thermal history around neutrino decoupling, the parametrization of the relativistic energy budget in terms of $N_{\rm eff}$, and the basic ingredients of Tsallis statistics and its generalized distribution functions, we are now in position to address the central aim of this work: to implement a controlled deformation of the neutrino phase-space energy density integral. In practice, we treat the neutrino sector as the only departure from the standard thermal picture, while keeping the electromagnetic plasma (photons and $e^\pm$ prior to annihilation) in its conventional form. This choice isolates the impact of nonextensivity on the relativistic content of the Universe and allows for a direct mapping between the Tsallis parameter $q$ and deviations in the neutrino energy density, and hence in $N_{\rm eff}$. In the following subsections we define the deformation scheme and the assumptions adopted, derive the corresponding expressions for the deformed thermodynamic integrals, and describe the inference strategy and datasets used to obtain bounds on $q$.

\subsection{Deformation scheme and assumptions}
We implement the deformation of the neutrino energy density within Tsallis statistics by evaluating the standard phase-space expression in Eq.~\eqref{eq:densidad_energia} with the generalized distribution \eqref{eq:fq} appropriate for neutrinos (fermions, hence $\xi=+1$),
\begin{align}
  \rho_q \;=\; g \int \frac{d^3p}{(2\pi)^3}\,E(p)\,f_q(E)\,,
  \label{eq:densidad_energia_Q} 
\end{align}
thereby encoding the nonextensive effects directly at the level of the microscopic occupation numbers. In the ultra-relativistic regime it is convenient to introduce the dimensionless variable $z\equiv \beta E = E/T$ (we assume chemical potential $\mu = 0$) and to express the fermionic Tsallis distribution as
$f_q(z)=\bigl[e_q(z)+1\bigr]^{-1}$. One may then define the dimensionless moment ratio,
\begin{equation}
R_\rho^{(\xi = +1)}(q)\equiv \frac{\rho_q}{\rho^{\rm std}} = 
\frac{\displaystyle\int_{0}^{z_{\max}} \frac{z^{3}\,dz}{e_q(z)+1}}
{\displaystyle\int_{0}^{\infty} \frac{z^{3}\,dz}{e^{z}+1}},
\qquad
z_{\max}=\begin{cases}
\dfrac{1}{1-q}, & q<1,\\[4pt]
\infty, & q\ge 1,
\end{cases}
\label{eq:Rrho-exacto}
\end{equation}
which measures the departure from the Boltzmann-Gibbs result at fixed temperature. For $q<1$ the support of the distribution is compact, and the integral is naturally cut off at $z_{\max}=1/(1-q)$, whereas for $q\ge 1$ the integration extends to infinity. For $q>1$ the $q$-exponential exhibits a power-law tail,
$e_q(z)\sim z^{1/(q-1)}$ (up to a constant prefactor), so that
\begin{equation}
\frac{z^3}{e_q(z)+1}\sim z^{\,3-\frac{1}{q-1}}\qquad (z\to\infty).
\end{equation}
Therefore the integral converges only if $3-\frac{1}{q-1}<-1$, i.e.\ $q<5/4$ \cite{Ishihara2015TsallisSigma}. The purpose of introducing $R_\rho(q)$ is to factor out the standard ultra-relativistic scaling and obtain a simple and transparent mapping between the nonextensive parameter $q$ and an effective rescaling of the neutrino energy density, which will be directly translated into a deformed neutrino sector and, consequently, into shifts in $N_{\rm eff}$.

We quantify the impact of nonextensivity on the relativistic sector by introducing the shift in the effective number of neutrino species induced by the Tsallis deformation,
\begin{equation}
\Delta N_{\rm eff}(q)\;\equiv\;\widehat N_{\rm eff}(q)-N_{\rm eff}^{\rm std}
\;=\;\bigl[R_{\rho}^{\,(\xi=+1)}(q)-1\bigr]\,N_{\rm eff}^{\rm std}\,,
\label{DELTANEFF1}
\end{equation}
where $N_{\rm eff}^{\rm std}$ denotes the Standard-Model value and $R_{\rho}^{\,(\xi=+1)}(q)$ is the energy-density rescaling factor computed from the fermionic ($\xi=+1$) Tsallis distribution. In the scheme adopted here, the deformation is restricted to the neutrino sector, so that photons remain standard while neutrinos acquire an effective rescaling,
\begin{equation}
\rho_{\gamma,q}=\rho_{\gamma}\,,\qquad
\rho_{\nu,q}=R_{\rho}^{\,(\xi=+1)}(q)\,\rho_{\nu}^{\rm std}\,.
\end{equation}
Accordingly, after $e^\pm$ annihilation the total radiation energy density can be written as
\begin{equation}
\rho_{r,q}
\;=\;\rho_{\gamma}+\rho_{\nu,q}
\;=\;\rho_{\gamma}\Bigl[1+k\,\widehat N_{\rm eff}(q)\Bigr]
\;=\;\rho_{\gamma}\Bigl[1+k\,R_{\rho}^{\,(\xi=+1)}(q) N_{\rm eff}^{\rm std}\Bigr]\,,
\label{eq:rho_r_q}
\end{equation}
with $k=\frac{7}{8}\left(\frac{4}{11}\right)^{4/3}$ as defined in Eq.~\eqref{eq:factor} (note that we use the instant-decoupling factor as a first approximation for our purposes). Note that we do not modify the photon energy density, since photons play a central role in the recombination epoch, CMB radiation and provide the reference component in the standard definition of $N_{\rm eff}$. Equation~\eqref{eq:rho_r_q} therefore represents our effective radiation budget in the post-$e^\pm$ annihilation Universe when nonextensivity is assumed to affect only neutrinos through the Tsallis deformation. This one-parameter mapping between the microscopic deformation encoded in $q$ and the macroscopic shift in the relativistic energy density is the basis of our strategy to derive bounds on the Tsallis parameter.

\subsection{Inference methodology and datasets}
With the neutrino sector already deformed within Tsallis statistics and consistently embedded into the radiation content after $e^\pm$ annihilation, we now outline the statistical inference framework adopted in this work and summarize the datasets used to constrain $q$.

\begin{table}[!h]
  \centering
  \sisetup{
    separate-uncertainty = true,
    table-number-alignment = center,
    table-text-alignment = center
  }
  \caption{Summary of data used in the analysis.}
  \label{tab:exp-data}
  \begin{tabular}{@{} l l S[table-format=1.3(3)] l @{}}
    \toprule
    Observable & Symbol & {Value (1$\sigma$)} & Experiment/Dataset \\
    \midrule
    Radiation effective (CMB$+$BAO) & $N_{\rm eff}^{\rm CMB}$ &
      2.99 \pm 0.17 & Planck 2018 $+$ BAO \cite{Planck2018CosmoParams} \\
    Radiation effective (BBN) & $N_{\rm eff}^{\rm BBN}$ &
      2.88 \pm 0.16 & BBN (primordial abundances) \cite{CyburtFieldsOliveYeh2016} \\
    %\midrule
    %Theoretical reference (SM) & $N_{\rm eff}^{\rm std}$ &
      %3.0440 \pm 0.0002 &~ Neutrino decoupling \cite{Bennett:2021NeffII} \\
    \bottomrule
  \end{tabular}
\end{table}

The comparison between the model prediction and the measurements can be carried out via $\chi^2$ built from a combination of the observational uncertainties of BBN and CMB$+$BAO present in Table \ref{tab:exp-data} as follows
\begin{equation}
  \chi^2_{N_{\rm eff}}(q) = \chi^2_{\rm BBN}(q) + \chi^2_{\rm CMB+BAO}(q) 
  \;=\;
  \frac{\bigl[\widehat N_{\rm eff}(q)-\mu_{\rm BBN}\bigr]^2}{\sigma_{\rm BBN}^2}
  \;+\;
  \frac{\bigl[\widehat N_{\rm eff}(q)-\mu_{\rm CMB+BAO}\bigr]^2}{\sigma_{\rm CMB+BAO}^2},
  \label{eq:chi2_data_only}
\end{equation}
where $\mu$ corresponds to the mean and $\sigma$ the uncertainty of the observational values of BBN and CMB$+$BAO respectively. If we were to add a theoretical term through $N^{\rm std}_{\rm eff}$ it would act as a very strong attractor and force $q\to1$ nullifying the possibility of any small window of nonextensivity at the beginning of BBN, since we want to obtain a bound for $q$ that does not violate the observations. It is also useful to define a $\Delta \chi_{N_{\rm eff}}^2(q)$ as 
\begin{equation}
    \Delta \chi_{N_{\rm eff}}^2(q) = \chi^2_{N_{\rm eff}}(q) - \chi^2_{N_{\rm eff,min}}(q),\label{DELTAXINEFF}
\end{equation}
where $\chi^2_{N_{\rm eff,min}}(q)$ is the minimum value of \eqref{eq:chi2_data_only} and allows shifting the original function downward, making it possible to find where it touches zero.

\section{Results and Discussion}
\label{sec:results}

In this section we constrain the nonextensive parameter $q$ using the BBN and CMB$+$BAO determinations of $N_{\rm eff}$ summarized in Table~\ref{tab:exp-data}. 
We quantify the deformation of the radiation sector through the shift $\Delta N_{\rm eff}(q)$ induced by the rescaling $R_{\rho}^{(\xi=+1)}(q)$, assuming it affects only the neutrino background. 
We then construct $\chi^2_{N_{\rm eff}}(q)$ for each dataset and for their combination, and derive confidence intervals on $q$ at the 68\%, 95\%, and 99\% confidence levels. 
Although the CMB constrains $N_{\rm eff}$ at recombination, we interpret the resulting limits as bounds on the underlying deformation controlling the neutrino energy density and assume that the same $q$ applies over the thermal history relevant to BBN, so we quote the final constraint at $T\simeq 1~\mathrm{MeV}$.

\begin{figure}[!t]
  \centering
  \begin{minipage}{0.5\textwidth}
    \centering
    \includegraphics[width=\linewidth]{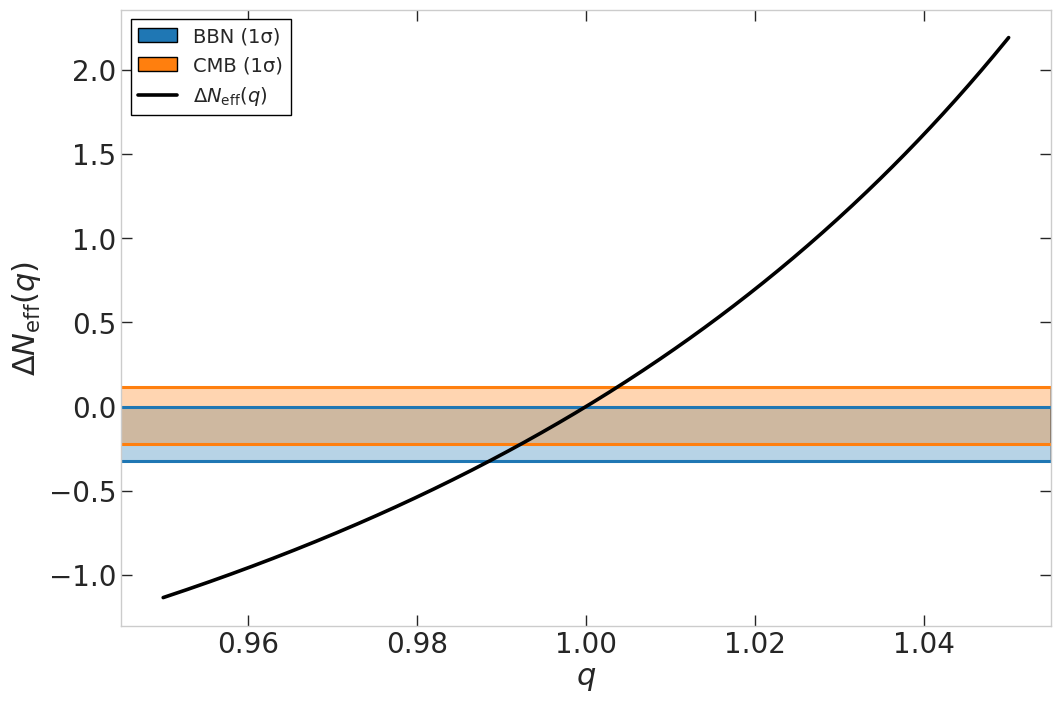}
  \end{minipage}\hfill
  \begin{minipage}{0.5\textwidth}
    \centering
    \includegraphics[width=\linewidth]{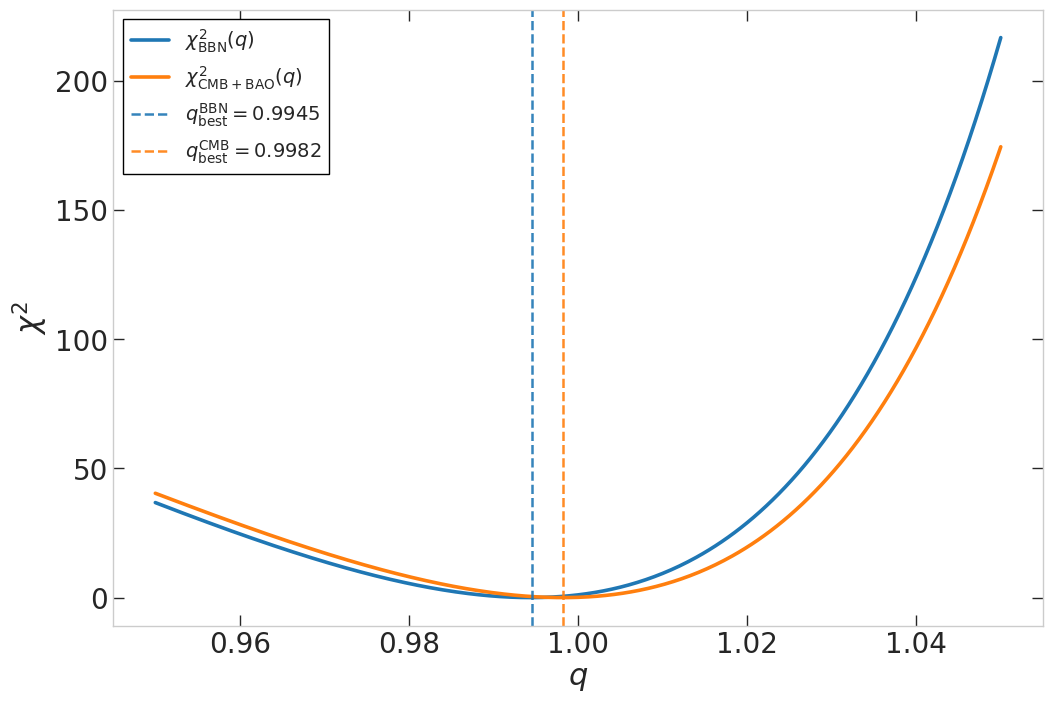}
  \end{minipage}
  \caption{\textbf{Left:} Non-extensive prediction for $\Delta N_{\rm eff}(q)$ induced by a neutrino-only deformation through the rescaling $R_{\rho}^{(\xi=+1)}(q)$, using $\Delta N_{\rm eff}(q)=\big(R_{\rho}^{(\xi=+1)}(q)-1\big)\,N_{\rm eff}^{\rm std}$. The horizontal bands show the $1\sigma$ regions of $N_{\rm eff}$ from BBN and CMB$+$BAO around the standard value; values of $q$ compatible with both datasets are those for which the curve lies within the overlap. \textbf{Right:} $\chi^2$ profiles as a function of $q$ for BBN and CMB$+$BAO, $\chi^2_{\rm BBN}(q)$ and $\chi^2_{\rm CMB+BAO}(q)$, with vertical lines indicating the corresponding best-fit minima $q_{\rm best}^{\rm BBN}$ and $q_{\rm best}^{\rm CMB}$.}
  \label{fig:deltaneffyseparado}
\end{figure}

\begin{figure}[!t]
  \centering
  \begin{minipage}{0.49\textwidth}
    \centering
    \includegraphics[width=\linewidth]{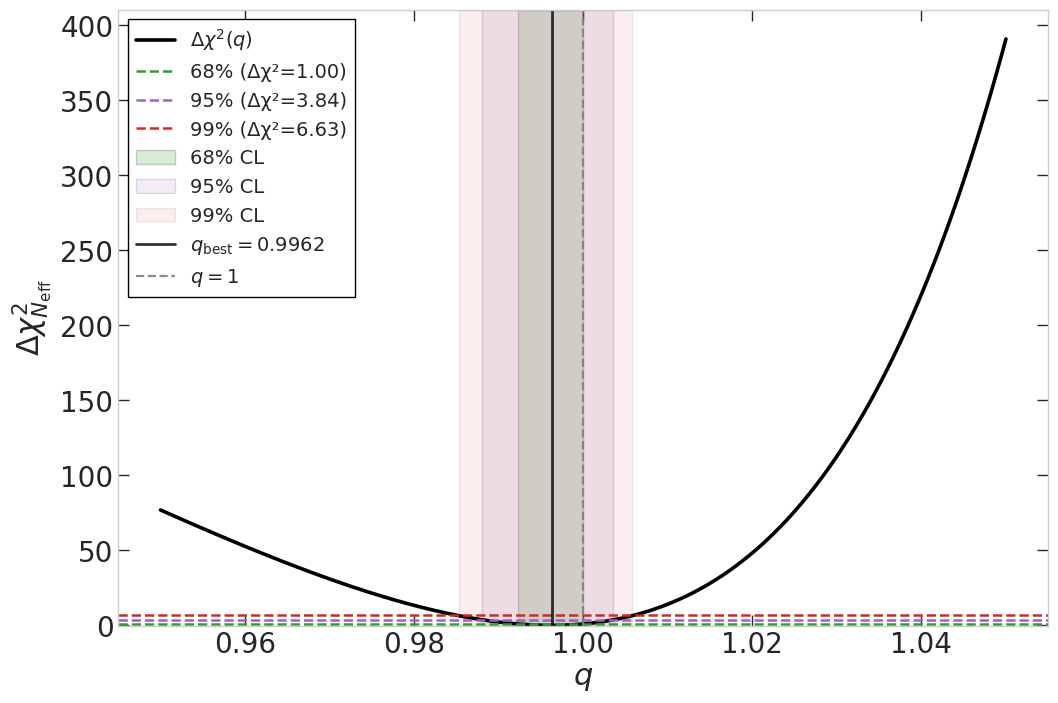}
  \end{minipage}\hfill
  \begin{minipage}{0.49\textwidth}
    \centering
    \includegraphics[width=\linewidth]{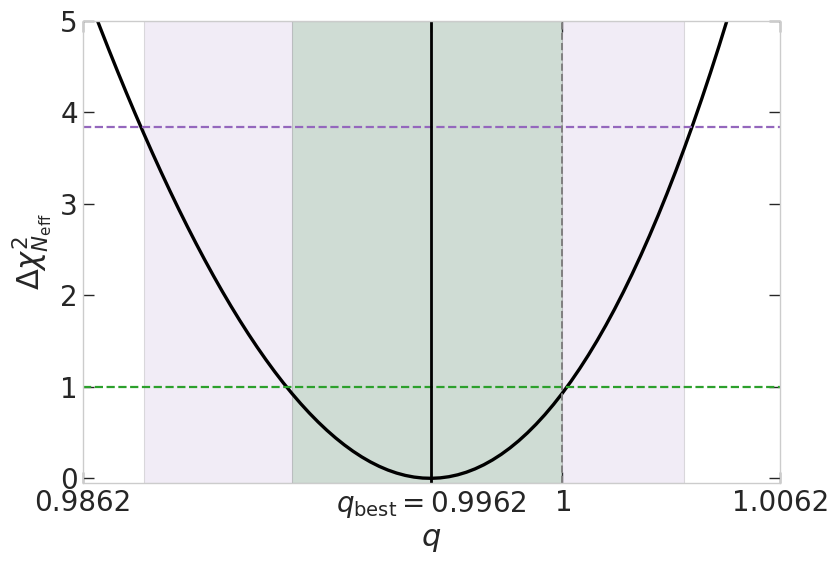}
  \end{minipage}
  \caption{\textbf{Left:} Profile likelihood for the combined constraint, $\Delta\chi^2_{N_{\rm eff}}(q)\equiv \chi^2_{N_{\rm eff}}(q)-\chi^2_{N_{\rm eff},{\rm min}}$, obtained under the neutrino-only rescaling scheme. Horizontal lines mark the 68\%, 95\%, and 99\% confidence levels (one effective parameter), and the shaded bands indicate the corresponding allowed intervals around $q_{\rm best}$; the gray line indicates $q=1$. \textbf{Right:} Zoom around the minimum highlighting the 68\% and 95\% confidence levels, with the black vertical line marking $q_{\rm best}$ and the gray dotted line indicating $q=1$.}
  \label{fig:deltayzoom}
\end{figure}
Figure~\ref{fig:deltaneffyseparado} shows the predicted shift $\Delta N_{\rm eff}(q)$ as a function of the Tsallis parameter $q$, together with the $1\sigma$ ranges from BBN and CMB$+$BAO. The observational bands restrict $q$ to lie close to the extensive limit, indicating that only percent-level departures from $q=1$ are compatible with the data in the neutrino-only deformation scheme. The monotonic behavior of $\Delta N_{\rm eff}(q)$ directly reflects the rescaling of the neutrino energy density encoded in $R_{\rho}^{(\xi=+1)}(q)$: values of $q$ that enhance (suppress) the relativistic tail lead to $\Delta N_{\rm eff}(q)>0$ ($<0$), and are accordingly bounded by the allowed ranges. The right panel displays the corresponding $\chi^2$ profiles for each dataset, $\chi^2_{\rm BBN}(q)$ and $\chi^2_{\rm CMB+BAO}(q)$, whose minima are both found near $q=1$, showing that the preferred values inferred from the two epochs are mutually consistent within uncertainties.

Figure~\ref{fig:deltayzoom} presents the combined constraint obtained from Eq.~\eqref{eq:chi2_data_only} in terms of $\Delta\chi^2_{N_{\rm eff}}(q)$ defined in Eq.~\eqref{DELTAXINEFF}, indicating the 68\%, 95\%, and 99\% confidence regions and the best-fit value $q_{\rm best}$. As expected, the joint fit narrows the allowed interval around $q_{\rm best}$ relative to the individual constraints, providing the most stringent bound on the departure from extensivity, $|q-1|$, in the temperature range relevant for BBN.

\FloatBarrier

\begin{table}[!h]
\centering
\small
\setlength{\tabcolsep}{4.5pt}
\renewcommand{\arraystretch}{1.15}
\caption{Constraints obtained for $q$ and the best-fit minimum for BBN and CMB$+$BAO.}
\label{tab:q-fits-neff}
\begin{tabular}{lcccc}
\toprule
Dataset & $q_{\rm best}$ & 68\% CL & 95\% CL & 99\% CL \\
\midrule
BBN        & \num{0.9945} & [\num{0.9888}, \num{0.9998}] & [\num{0.9825}, \num{1.005}] & [\num{0.9785}, \num{1.008}] \\
CMB+BAO    & \num{0.9983} & [\num{0.9925}, \num{1.004}] & [\num{0.9863}, \num{1.009}] & [\num{0.9820}, \num{1.012}] \\
Combined   & \num{0.9962} & [\num{0.9927}, \num{1.000}] & [\num{0.9891}, \num{1.004}] & [\num{0.9868}, \num{1.006}] \\
\bottomrule
\end{tabular}
\end{table}
\FloatBarrier

 Taking the biggest deviation from $q_{\rm best}$ from the combined $\chi^2_{N_{\rm eff}}(q)$ analysis in Table \ref{tab:q-fits-neff} we have
\begin{equation}
    |q-1| \le 1.09\times 10^{-2}\quad (95\%~\mathrm{CL}),\label{cota95}
\end{equation}
\begin{equation}
    |q-1| \le 1.32\times 10^{-2}\quad (99\%~\mathrm{CL}),\label{cota99}
\end{equation}
which are our final constraints for the nonextensive parameter $q$ at $T\simeq1$ MeV. Table \ref{tab:q-fits-neff} summarizes the intervals with respect to the Confidence Levels for the non-extensivity parameter $q$ via an upper and lower bound and a best fit given by $q_{\rm best}$ from a $\chi^2$-type fit combining the CMB$+$BAO and BBN data of Table \ref{tab:exp-data}. These results are consistent with the extensive limit $q=1$ and constrain any neutrino-sector nonextensive deformation to the percent level during the epoch relevant for BBN.

\section{Conclusions and outlook}
\label{sec:conclusion}
In this work we investigated a controlled deformation of the neutrino radiation sector in the early Universe within the framework of Tsallis statistics. Starting from the Curado-Tsallis prescription, we introduced the generalized equilibrium distribution functions and evaluated the corresponding energy-density integral for ultra-relativistic fermions. This allowed us to define the rescaling factor $R_{\rho}^{(\xi=+1)}(q)$, which provides a direct mapping between the microscopic deformation encoded in the Tsallis parameter $q$ and the macroscopic radiation content parametrized by the effective number of neutrino species. In the neutrino-only scheme adopted here, this mapping translates into a shift $\Delta N_{\rm eff}(q)=\bigl[R_{\rho}^{(\xi=+1)}(q)-1\bigr]\,N_{\rm eff}^{\rm std}$ while keeping the photon sector unmodified.

We confronted this prediction with the current BBN and CMB$+$BAO determinations of $N_{\rm eff}$ by constructing $\chi^2_{N_{\rm eff}}(q)$ for each dataset and for their combination. The resulting likelihood profiles prefer values close to the extensive limit and yield percent-level bounds on departures from $q=1$. From the combined analysis we obtained the constraints
$|q-1|\le 1.09\times 10^{-2}$ (95\% CL) and $|q-1|\le 1.32\times 10^{-2}$ (99\% CL),
which we interpret as limits on the underlying deformation controlling the neutrino energy density and quote at $T\simeq 1~\mathrm{MeV}$, relevant for the onset of BBN. Our limits should be understood within the minimal setup adopted here, namely a neutrino-only deformation with vanishing chemical potential and a constant $q$ over the relevant thermal history.

These results show that any nonextensive modification of the neutrino background compatible with present measurements must be small, providing a quantitative benchmark for scenarios in which mild departures from Boltzmann-Gibbs statistics arise from long-range interactions or residual correlations in the primordial plasma. Several extensions of this work are straightforward. On the theory side, one may relax the assumption of a constant $q$ and consider a temperature-dependent deformation $q(T)$, or explore alternative constraint prescriptions within nonextensive thermodynamics. On the phenomenology side, the same strategy can be applied to other early-Universe observables sensitive to the relativistic energy density, including future CMB measurements with improved sensitivity to $N_{\rm eff}$, which will further sharpen the allowed window around the extensive limit.

\acknowledgments

MPG. acknowledges Vicerrectoría de Investigación y Desarrollo Tecnológico (VRIDT) at Universidad Católica del Norte (UCN) for the scientific support provided by Núcleo de Investigación en Simetrías y la Estructura del Universo (NISEU-UCN), Resolución VRIDT N°200/2025.

MPG. acknowledges to my fellow roommates of the graduate program at \textit{Universidad Católica del Norte}.

MPG. acknowledges the financial support of the \textit{Dirección general de postgrado}.

\end{document}